\documentclass{article}

\usepackage{PRIMEarxiv}
\usepackage{float} 
\usepackage[utf8]{inputenc} 
\usepackage[T1]{fontenc}    
\usepackage{hyperref}       
\usepackage{url}            
\usepackage{booktabs}       
\usepackage{amsfonts}       
\usepackage{nicefrac}       
\usepackage{microtype}      
\usepackage{lipsum}
\usepackage{fancyhdr}       
\usepackage{graphicx}       
\graphicspath{{media/}}     

\title{Allen: Rethinking MAS Design through Step-Level Policy Autonomy}

\author{
 Qiangong Zhou \\
  \texttt{zhouqg@motern.com} \\
  \And
 Zhiting Wang \thanks{Corresponding author}\\
  \texttt{wangzt@motern.com} \\
  \And
 Mingyou Yao \\
  \texttt{yaomy@motern.com} \\
  \And
 Zongyang Liu \\
  \texttt{lzy@motern.com} \\
  \And
  Shenzhen Motern Technology Co., Ltd.
}

\begin{document}
\maketitle

\begin{abstract}
We introduce a new Multi-Agent System (MAS) - Allen, designed to address two core challenges in current MAS design:
(1) improve system's policy autonomy, empowering agents to dynamically adapt their behavioral strategies, and (2) achieving the trade-off between collaborative efficiency, task supervision, and human oversight in complex network topologies. 

Our core insight is to redefine the basic execution unit in the MAS, allowing agents to autonomously form different patterns by combining these units.
We have constructed a four-tier state architecture (Task, Stage, Agent, Step) to constrain system behavior from both task-oriented and execution-oriented perspectives. 
This achieves a unification of topological optimization and controllable progress.

Allen grants unprecedented Policy Autonomy, while making a trade-off for the controllability of the collaborative structure. 
The project code has been open source at: \url{https://github.com/motern88/Allen}

\end{abstract}


\section{Introduction}
Multi-Agent Systems (MAS) can handle complex parallel tasks better than single-agent systems, attracting widespread attention in recent years.
When designing a MAS, researchers often break it down into two sub-problems: 
\begin{enumerate}
\item \textbf{Intra-Agent Architecture Design:} How to design an agent's internal logic/workflow.
\item \textbf{Inter-Agent Collaboration Framework Design:} How to design the collaboration framework between agents.
\end{enumerate}

For the first sub-problem, how to design an agent's internal logic/workflow, a key 2024 survey paper\cite{1} summarized various agent architectures.
These were categorized into 18 distinct patterns, along with guidance on selecting the right pattern for different scenarios.
These patterns include: prompt optimization and response generation, Retrieval-Augmented Generation (RAG), single-path and multi-path planning, self-reflection mechanisms, voting-based multi-agent collaboration, role-based collaborative frameworks, and debate-driven collective decision-making paradigms.

Currently, it has become extremely difficult to propose fundamental innovations at the level of agent behavior patterns - that is, to design completely novel behavior patterns for specific new scenarios.
The good news is that we believe that the current bottleneck for agents lies not in proposing more new patterns, but rather in making proper decisions to enable agents to use the most suitable behavior patterns - including specific workflows or task procedures - across different scenarios.

We argue that an agent's operational logic and behavior patterns should not be confined to a single, fixed paradigm. 
Instead, it should possess the capability to dynamically determine and select optimal patterns based on the situational context.
We define the scope and flexibility of such decision-making capacity as the system's \textbf{Policy Autonomy}, whose upper bound determines the system's intelligence and adaptability levels.

For instance, in some systems, agents are strictly constrained within predefined workflows, where their Policy Autonomy is limited to making choices at a few predetermined branch points.
In other systems, Policy Autonomy manifests through the system's ability to autonomously select and combine multiple predefined agents with specialized functions according to task requirements.

Therefore, we redefine the first sub-goal — "designing reasonable internal agent logic/workflows" — as a more fundamental question:
\begin{itemize}
\item How to maximize a system's Policy Autonomy?
\end{itemize}

Concerning the second sub-problem—constructing the collaborative framework among agents—a paper\cite{2} published by Google in February 2025 states that the key to enhancing the performance of multi-agent frameworks lies in optimizing the topological structure. 
Our methodology aligns with this principle.
First, we optimize the internal decision-making topology within each agent.
Second, at the system level, we unlock the potential for optimizing the overall topology by empowering agents with the autonomy to define task workflows and determine communication timing with other agents.

Optimizing the topological structure of Multi-Agent Systems (MAS) can improve collaborative efficiency among agents.
However, in practical applications, we must also consider two additional usability factors: task progress monitoring and human intervention capability.
Yet, complex and flexible topological structures often conflict with both progress monitoring and human intervention.
Therefore, we reformulate the second objective - designing the inter-agent collaborative task framework - as:
\begin{itemize}
\item How can Multi-Agent Systems achieve optimal balance between collaborative efficiency, progress monitoring, and human intervention capability?
\end{itemize}

To achieve these two primary objectives, we redefine the minimal unit of agent execution and introduce a four-tier state hierarchy comprising Task, Stage, Agent, and Step.
To this end, we have designed and implemented a Multi-Agent System named Allen.

Specifically, the contributions of this paper are as follows:
\begin{itemize}
\item \textbf{Novel Perspective}. 
We introduce "System Policy Autonomy" as a novel perspective to guide the design and evaluation of agent systems.
\item \textbf{Execution Model}. 
We propose a Step-centric execution model to reconstruct the agent's operational logic. 
This model achieves a state-of-the-art level of System Policy Autonomy, enabling it to simulate and execute virtually any existing agent work pattern.
\item \textbf{Collaborative Architecture}. 
Building upon this execution model, we have constructed a four-tier collaborative architecture for the multi-agent system.
This architecture strikes an exceptional balance among synergistic efficiency, progress observability, and human intervenability.
\end{itemize}

\section{Related Work}

\begin{figure}[H]
  \centering
  \includegraphics[width=1.0\textwidth]{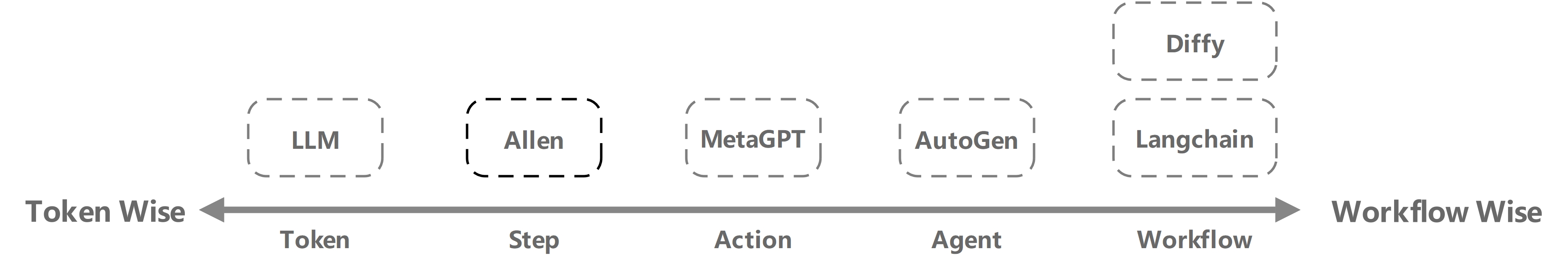}
  \caption{
  We evaluate different agent frameworks from the perspective of Policy Autonomy.
  The two ends of this spectrum represent two extreme paradigms: 
  on the left, the token-wise approach treats a single token as the minimal unit of execution, thus offering the highest degree of Policy Autonomy.
  On the right, the workflow-wise approach uses a predefined workflow as the minimal unit of execution, resulting in the lowest Policy Autonomy.
  Our architecture, Allen, is positioned within this spectrum at the step-wise level, situated closer to the left side.
  }
  \label{figure1}
\end{figure}

If we examine the agent architectures from the perspective of 'Policy Autonomy,' a non-rigorous conceptual comparison is shown as Figure \ref{figure1}.

All current mainstream agent implementation frameworks  (not the final products) can all be evaluated along the Policy Autonomy spectrum shown in Figure \ref{figure1}.
On the far left of the spectrum, representing the highest policy autonomy, are foundational LLMs, where every generated token is effectively an autonomous decision.
On the far right, representing the lowest policy autonomy, are workflow-level systems.
With these, we can only choose between pre-defined workflows and must build a new, complete, and adapted workflow for each new task and environment.

In Figure \ref{figure1}, frameworks like LangChain\cite{3} and Dify\cite{4} implement Policy Autonomy at the workflow level, requiring custom workflow construction for each scenario.
Strictly speaking, these are workflow implementation tools.
In contrast, AutoGen\cite{5} achieves agent-wise Policy Autonomy - users only need to create new agents for novel scenarios, and the framework autonomously selects participating agents.

MetaGPT\cite{6} demonstrates even higher Policy Autonomy, achieving action-wise decision capability. 
Here, each agent can dynamically select from predefined actions during execution, where actions closely resemble agent behavioral patterns.
Agents operate via think-act cycles, autonomously choosing different actions in each iteration from a skill and tool library.
Thus, MetaGPT\cite{6} represents a highly flexible framework with superior Policy Autonomy.

However, we aim to avoid implementing task-specific actions when adapting to new scenarios.
We believe action-level composition (using predefined skills/tools) doesn't represent the theoretical limit of agent Policy Autonomy.
Instead, we pursue minimal executable units - when properly sequenced, these units should macroscopically generate diverse agent behaviors without manual logic adaptation.
Natural transitions between units should enable autonomous workflow determination.

We define the minimal execution unit within our system as a "Step". 
Consequently, our Allen framework implements a system with step-wise policy autonomy.
We posit that a purely token-wise system (such as a raw LLM), despite having the highest theoretical freedom, is impractical for complex tasks due to its lack of structured task-orientation and reliable state judgment.
Furthermore, imposing fixed output validation on it is tantamount to degrading it to a workflow-wise system.

Therefore, to our knowledge, Allen is the agent system that achieves the highest degree of policy autonomy while maintaining a crucial balance between structure and flexibility.
It empowers each agent to autonomously select and combine Steps, enabling it to dynamically generate and execute a sequence of Actions best suited for the current context.
This achieves adaptive behavioral patterns without the need for manual pre-define.

\section{Agent's Internal Mechanisms}
In this chapter, we will introduce how we construct a coherent internal operational logic/pattern for an Agent.
In the Introduction of Chapter 1, we reframed this objective as a question of "how to enhance the Policy Autonomy of a single-agent system."

A natural question that arises is: how can the various agent operational patterns (e.g., reflection, planning) summarized in the survey\cite{1} be dynamically integrated into a single agent's execution flow?
In other words, how can an agent autonomously decide when to reflect, when to plan, and when to call tools?
We believe it must achieve two signature capabilities:
\begin{enumerate}
\item The Agent can determine which tools and skills to use.
\item The Agent can determine its own operational logic.
\end{enumerate}

From the perspective of Policy Autonomy, the higher the system's overall policy autonomy, the more readily it can achieve the two aforementioned hallmark capabilities.
The ability for an Agent to decide which tools and skills to use requires that our Policy Autonomy operates at the tool and skill level.
The ability for an Agent to determine its own operational logic requires that this logic be dynamic.
A natural synthesis, therefore, is for the Agent to autonomously select individual tools and skills, which in turn form a complete and dynamic operational logic.
This approach enables the Agent to define and modify its own operational logic. 

So, the problem is refined to implementing an Agent execution mechanism that makes decisions at the tool and skill level.
Furthermore, each decision-making unit must be sufficiently small, allowing macroscopic internal operational patterns to be assembled from these small, individual decisions.

We define the minimal unit of our Agent's execution as a "Step", which also serves as our system's smallest decision-making unit.
Certain Steps can determine what the next Step will be. 
This way, we only need to implement a set of fundamental Steps and then leave it to the LLM to decide when to use each specific one.
This creates a highly extensible and dynamic operational logic for the Agent that can adapt to any task.
At this point, our Agent is equipped with both of the key capabilities mentioned earlier:
(1) The Agent can decide which tools and skills to use.
(2) The Agent can determine its own operational logic.

From a topological standpoint, our Steps exhibit relational dependencies.
Certain Steps—such as those for Planning, Reflection, and Decision—possess the ability to append new Steps to the Agent's execution queue.
Therefore, within our collection of all possible Steps, these can be seen as "pointing to" other Steps.
An Agent's execution can "jump" from one of these Steps to another—for instance, a Planning Step might schedule several new Steps.
Critically, these newly scheduled Steps can, in turn, re-engage Steps that have already been executed, particularly those with decision-making and planning capabilities.
Consequently, within the entire space composed of all Steps, the directional relationships between them form multiple cyclical structures (loops).
This ensures the Agent has the capability for autonomous and continuous execution, without a human needing to periodically "wind it up".

\subsection{Execution Process of Step}
\label{sec:3.1}
After introducing our use of Steps to define the minimal execution units, we now explain how the system's fundamental unit - Step - is executed.
There are two types of Steps, skills and tools.
We classify all steps requiring LLM calls as Skills, and all steps not involving LLM calls as Tools. 
The implementations of Skills and Tools are detailed in  \ref{app:a}.

The execution of a Step may involve either a Skill call or a Tool call, determined by the executor information specified in that Step.
The concrete execution flow of Steps is shown in the following diagram:
\begin{figure}[H]
  \centering
  \includegraphics[width=0.55\textwidth]{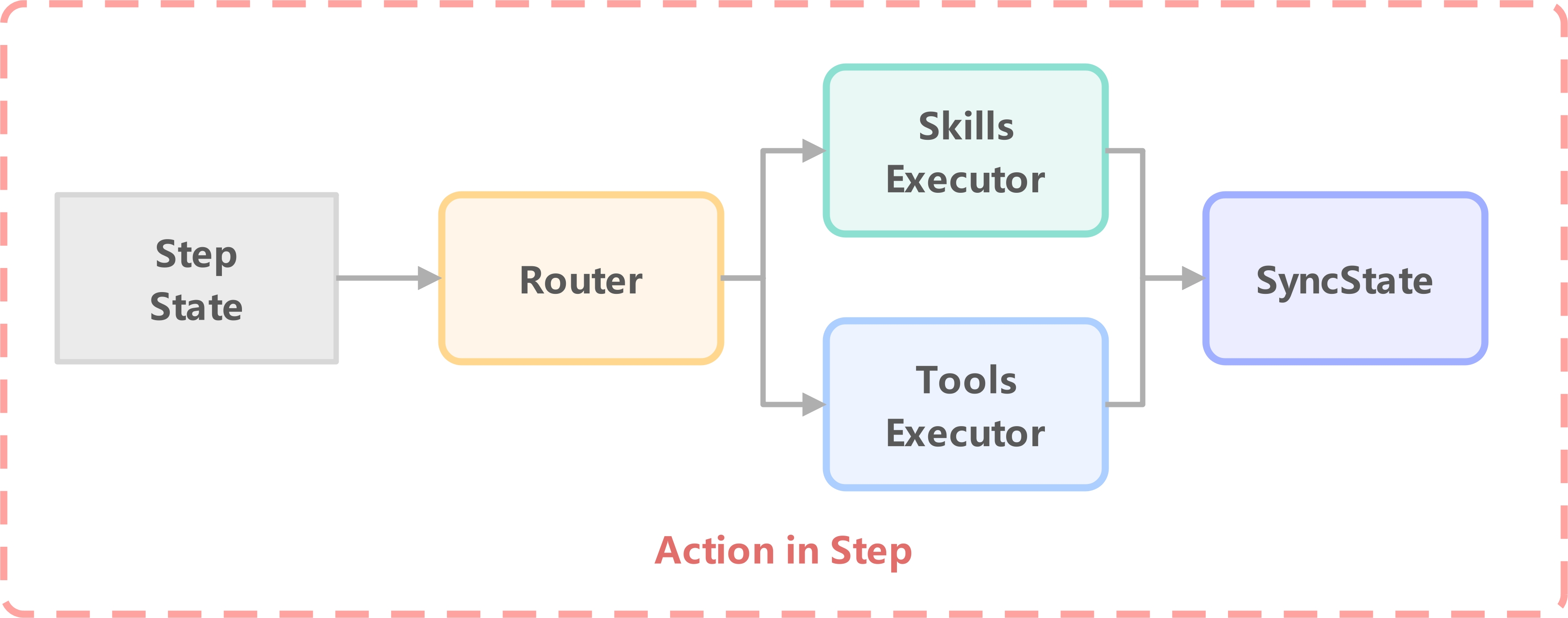}
  \caption{
  The Execution Flow of a Step.
  The process is orchestrated around the Step State, which encapsulates all state information required for execution.
  When a Step is initiated:
  (1) The Router component analyzes the Step State to determine the type of Step and invokes the corresponding Executor.
  (2) The designated Skill/Tool Executor performs the specific interaction (e.g., communicating with the LLM or calling an external API) based on the context provided by the Step State.
  (3) Upon completion, the Executor generates additional instructions to guide the SyncState component in performing task-level state synchronization and updates.
  }
  \label{figure2}
\end{figure}

In implementation, we instantiate a Step State for each concrete Step to carry detailed execution information including the Step's unique identifier, type, and intent.
We define the execution of a single concrete Step as one Action of an Agent.
In other frameworks like MetaGPT\cite{6}, an Action is defined as the execution of multiple concrete steps, thus requiring predefined Action types, whereas we only need to preset Executors for different Steps.

As shown in Figure \ref{figure2}, during each Action of an Agent, the current Step to be executed is processed.
The Agent first passes the executor information recorded in Step State to the Router, which then dispatches the Step State to the corresponding Executor instance.
For Skill-type Executors, the Agent updates both its own state and the Step State within the Executor;
for Tool-type Executors, the Agent performs actual interactions with the external environment and updates the Step State within the Executor.

When synchronization of non-local states (e.g., task information or other Agent states) is required, it is implemented through the global SyncState.
Upon completion, the Executor returns instructions to guide SyncState in performing global state updates.
The SyncState is explicitly invoked at the end of an Agent's Action to mark its completion.

This design isolates the execution of each Step.
We implement all basic Skill or Tool executors such that during each Step, the executor performs operations solely based on the content of Step State.
Certain decision-making skills (e.g., Planning, Reflection, Decision) empower Agents to modify their own workflows by appending new Step State to the execution queue.

\subsection{The Action of Agent}
\label{sec:3.2}
We have explained how a single Step works. 
Now, we will describe how an Agent executes a sequence of these Steps.

\begin{figure}[H]
  \centering
  \includegraphics[width=0.8\textwidth]{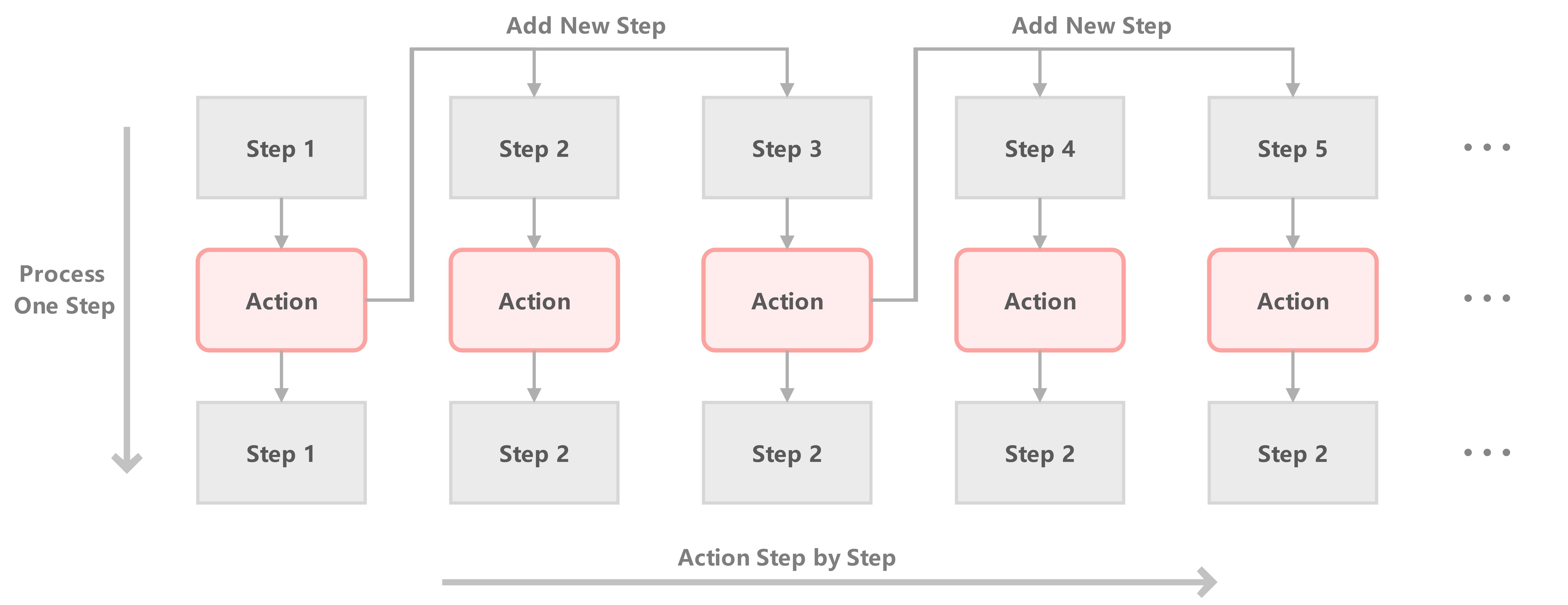}
  \caption{
  A flowchart showing the Agent's sequential execution of Steps.
  The horizontal axis is the timeline of Steps.
  The vertical axis is the process for each Step.
  Critically, some  decision-making Steps (like 1 and 3) can add new Steps to the queue.
  This allows the Agent to autonomously decide its own future work.
  }
  \label{figure3}
\end{figure}

An Agent works through a queue of pending Steps in order.
Figure \ref{figure3} shows this process.
The horizontal axis shows the order of execution from left to right.
The vertical axis shows the internal processing flow for a single Step.
During an Action (details refer to Figure \ref{figure2}, the system takes a Step State, executes the corresponding operation, and updates the Step State with the outcome.
The key feature of our model is that some decision-making Steps can add *new* Steps to the Agent's queue (exemplified by Step 1 and Step 3 in Figure \ref{figure3}).
When such a Step is executed, its resulting Action includes appending one or more new Step State objects to the end of the pending queue.
This is how the Agent dynamically controls its own behavior and autonomously decides what to do next.

We find this architecture similar to RNNs\cite{7}.
In terms of dynamic decision-making and flexible adaptation, RNNs\cite{7} make decisions based on current input and the previous hidden state to determine the next timestep's output.
Thus, RNNs\cite{7} possess the capability at each timestep to make different decisions based on historical information.
Similarly, our Agent dynamically adjusts subsequent workflows based on the current Step's state and decisions, thereby influencing the entire execution path.
This flexibility parallels how RNNs dynamically generate next-step outputs based on historical states and current inputs.

A key function of an RNN is to propagate its hidden state across timesteps, so it can capture information and model long-term dependencies within a sequence.
In our architecture, the Agent State plays a similar role.
The Agent State functions as a persistent repository of the agent's knowledge and context. 
This mechanism is essential for maintaining contextual awareness and resolving long-term dependencies.

\subsection{Agent State}
\label{sec:3.3}
The Agent State is the core data structure for an agent's runtime status. It records the agent's inherent attributes and the persistent memory (Appendix \ref{app:b}) accumulated during its execution.
This persistent memory is managed autonomously by the agent; its contents are continuously appended and are not automatically reset upon task completion unless explicitly cleared by the agent.
This mechanism ensures that the agent can effectively maintain long-range contextual dependencies when handling complex, long-horizon tasks.

All agents are instances of the same class, distinguished only by their individual Agent State.
This State encapsulates all information required to define an agent, allowing it to be perfectly replicated or restored.
Specifically, the State defines the agent’s role, personality, and its access permissions for tools and skills.

This architecture significantly simplifies building multi-agent systems.
Developers no longer need to write custom code for each agent; they only need to define its Agent State in a configuration file.
Furthermore, this grants agents the ability to dynamically create new, non-predefined agents, provided that the manager Agent can generate the corresponding configuration file.

\section{Inter-Agent Collaboration Mechanism}
In the previous chapter, we detailed the internal mechanics of a single agent.
We now shift our focus to how multiple agents can collaborate to accomplish complex tasks.

We posit that an effective multi-agent collaboration framework must ensure the task progress is clearly traceable, yet achieving this presents a significant challenge.
As noted in the survey\cite{1}, the accountability process in multi-agent collaboration is exceedingly complex.
This complexity stems from the intricate interactions within the ecosystem among various stakeholders, agents based on foundation models, non-agent AI models, and conventional software applications.
Highly autonomous agents can delegate tasks or even create other agents or tools to execute sub-tasks.
In such scenarios, responsibility and accountability become deeply intertwined among multiple entities.

\begin{figure}[H]
  \centering
  \includegraphics[width=1.0\textwidth]{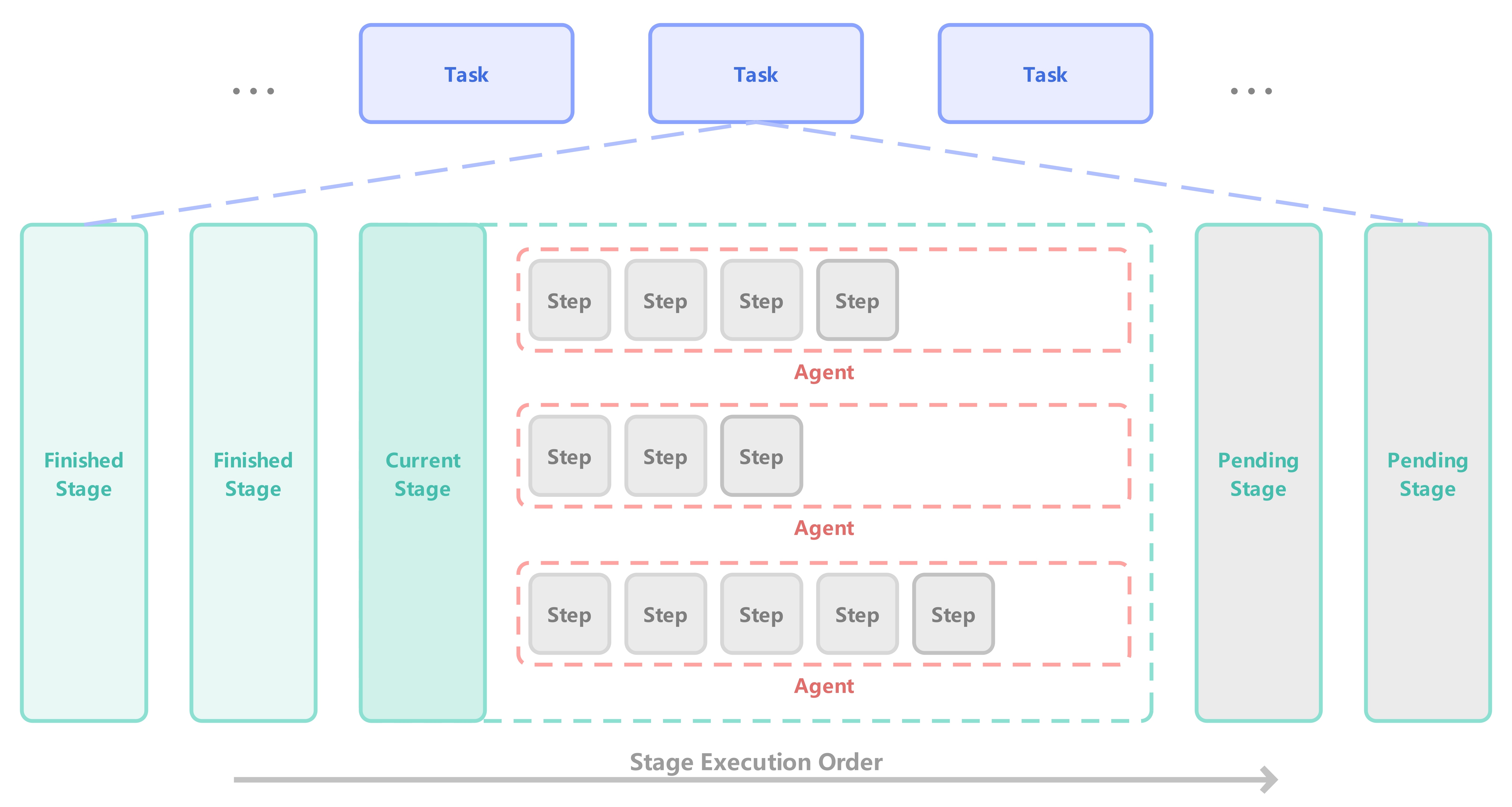}
  \caption{
  Our proposed Multi-Agent System (MAS) architecture.
  It uses four hierarchical state levels—Task, Stage, Agent, and Step—to record and track the task process across the system.
  The execution flow follows this hierarchy:
   (1) Within the MAS, multiple Tasks are executed in parallel, Stages are executed sequentially in each Task; 
   (2) Within each Stage, multiple Agents operate in parallel, Steps are executed sequentially in each Agent.
  }
  \label{figure4}
\end{figure}

To achieve this, we have implemented an architecture as shown in Figure \ref{figure4}, which systematically manages task execution through two task-level states (Task State and Stage State) and two agent execution-level states (Agent State and Step State).

We use Agent and Step levels to ensure that each Agent has strong Policy Autonomy, guaranteeing robust problem-solving capability for individual tasks.

We use Task and Stage levels to ensure clear task division.
The sequential execution between Stages and the parallel execution of multiple Agents within a Stage can adapt to most complex task frameworks.

\subsection{Status Records Task Information}
We will now detail how the task process is tracked and managed via this state-based system.
As seen in Figure \ref{figure4}, this hierarchical structure decomposes a Task into multiple Stages that are executed sequentially, ensuring only one Stage is active at any given time.
Within each Stage, multiple Agents operate in parallel to achieve the stage's overall objective; each Agent, in turn, completes its assigned sub-goal by executing a sequence of specific Steps (with internal mechanics detailed in Figure \ref{figure3}.

We first briefly introduce the actual operational logic corresponding to these four states:

\vspace{1em}
\textbf{Task State}

When the system is assigned multiple tasks, each task creates its own task group.
A task group consists of multiple Agents working on that task.
An Agent may participate in multiple task groups simultaneously, but a task group focuses on only one task.
Therefore, the Task State records all Agents in a task group and the complete task flow, including all Stages under that task.

When the task is completed, the task group disbands, and the Task State is cleared.

\vspace{1em}
\textbf{Stage State}

For complex tasks, an Agent often breaks the task into multiple sub-goals executed sequentially, with each Stage handling one sub-goal.
For more complex stage objectives, multiple Agents may execute them in parallel.
Agents can communicate to collaborate on achieving the stage goal.

The Stage State records detailed information about the sub-goal, including the specific objectives of each participating Agent.
The Stage State is not cleared immediately after completion but is retained until the entire Task ends.
This is because we need to track stage completion to determine Task completion.

\vspace{1em}
\textbf{Agent State}

As described in Section \ref{sec:3.3}, the Agent State is the core container for an Agent.
All information after Agent instantiation is stored here, including the Agent’s role background, skill and tool permissions and persistent memory.

The Agent State is primarily updated by the Agent itself, such as modifying its Agent Step to enable action and decision-making.
However, in some cases, manager Agents or human operators may modify another Agent’s state.
The Agent State is only cleared when the corresponding Agent is deactivated.

\vspace{1em}
\textbf{Step State}

As explained in Sections \ref{sec:3.1} and \ref{sec:3.2}, a Step is the smallest executable unit in the Multi-Agent System and the foundation of Agent activity.
Agents interact with the environment by executing individual Steps.

The Step State records details of each Step, including step type, executor name, step intent, and execution result.
The Step State is instantiated within the Agent State and managed by the Agent Step instance.
Step creation is determined by the Agent’s decision-making skills, while clearance is task-dependent (Agents are notified to release Steps related to a stage when it ends).

\begin{figure}[H]
  \centering
  \includegraphics[width=0.9\textwidth]{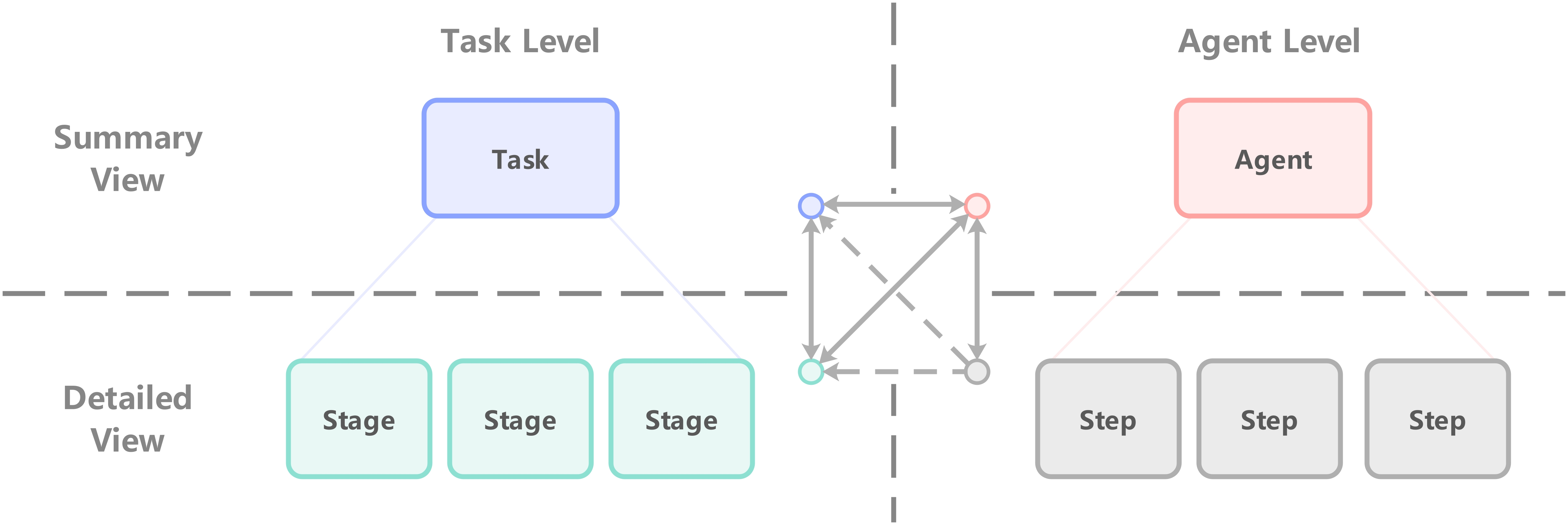}
  \caption{
  The relationship diagram of four states in Allen.
   On the left side, Task and Stage belong to task-level state recording; on the right side, Agent and Step belong to execution/Agent-level state recording.
   At the top, Task and Agent record global information from an overview perspective; at the bottom, Stage and Step record local detailed information.
   \\
   The arrows show how the four states relate.
   An arrow between two states means the starting state references the ending state.
   Bidirectional: solid lines; unidirectional: dashed lines.
  }
  \label{figure5}
\end{figure}

In addition, these four states also have interconnected properties, as shown in Figure \ref{figure5}.
Both Task State and Agent State record large amounts of global information, while Stage State and Step State only focus on small amounts of local details.
All Stages under a Task execute sequentially, and all Steps under an Agent also execute sequentially.

The arrow diagram at the center of the dividing line in Figure \ref{figure5} illustrates the cross-referencing relationships between the four states.
Dashed lines represent one-way references, and solid lines represent two-way references.
As shown:
\begin{itemize}
\item A Task needs to record its related Stages and Agents.
\item A Stage also records its related Task and Agent.
\item An Agent needs to record all related Tasks, Stages, and Steps.
\item A Step also needs to record all related Tasks, Stages, and Agents.
\end{itemize}

Between Task, Stage, and Agent, each state must record references to the other two states it is associated with.
This provides strong task-tracking capabilities.
However, Stage and Task do not point to specific Steps (they do not track which Step states belong to them)—they only point to the executing Agent (tracking which Agents are assigned).
This is because Step State is generated internally by the Agent and is only recorded in Agent State.

However, Step State points to Task and Stage to indicate which task and stage it belongs to.
This is necessary because an Agent may be assigned multiple Stages from different tasks and needs to distinguish which task/stage its current Step corresponds to.

From an information flow perspective, the implementation of these four states is highly beneficial for task decomposition and tracking.
At the task level, Task State and Stage State effectively record real-time task status while maintaining global (Task State) and local (Stage State) synchronization across Agents.
Meanwhile, Step State within an Agent regulates and standardizes the execution flow, serving as one source of the Agent's self-observation.

\subsection{Messages Carry Agent Communications}

\begin{figure}[H]
  \centering
  \includegraphics[width=1.0\textwidth]{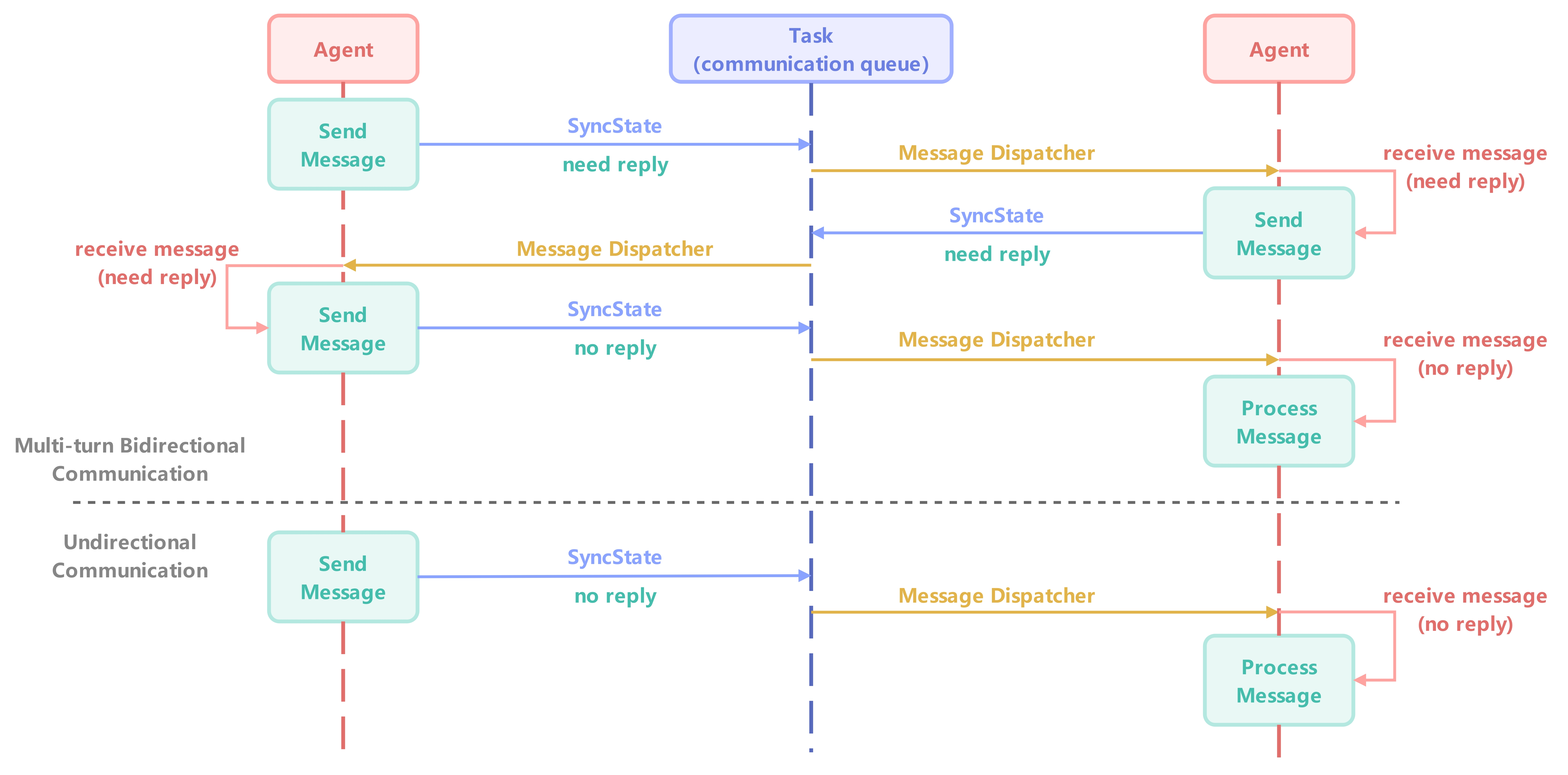}
  \caption{
  Agent communication process diagram.
  The upper part shows two-way multi-turn communication between two Agents; The lower part shows one-way single communication between two Agents.
  \\
  Agent communication starts with the Send Message skill.
  The message is passed to the recipient Agent through SyncState and Message Dispatcher.
  If the recipient replies, it triggers another Send Message; If there is no reply, it triggers Process Message instead.
  }
  \label{figure6}
\end{figure}

In the previous section, we used four states to maintain dynamic recording and tracking of task content.
In this section, we describe how we implement interactions in MAS — the communication mechanism between Agents.

We only need to implement message sending and the corresponding message receiving to support all communication methods in MAS, including: one-way messaging, two-way conversational communication, and broadcast messaging.

First, we simplify all messages in the MAS system into point-to-point communication between two Agents.
If it's a broadcast message, it can also be split into multiple one-to-one Agent communications.
Second, for point-to-point communication between two Agents, we combine two common approaches: 
long-term connections (for sustained conversations), and short-term one-way messaging (both now use the same communication method).

Following the design principle from earlier — using the smallest executable unit (Step) to build all MAS mechanisms — we implement any form of point-to-point communication with two specific skill Steps:
\begin{itemize}
    \item \textbf{Send Message:} \\
    The Agent organizes the message it wants to send, combining historical data (including recently received messages). It then packages the message into a specific structure.
    \item \textbf{Process Message:} \\
    The Agent only processes incoming messages in this Step.
\end{itemize}

As shown in Figure \ref{figure6}, the sender Agent uses its Send Message skill to organize and generate a message with a specific structure.
This message includes a property indicating whether a reply is needed. 
For one-way, single-time messages that only serve as notifications, this property will be marked as no reply required.

Since Agents cannot directly access each other (each Agent instance in the MAS is isolated), all messages are relayed through the Task layer.
The message generated by Send Message is added to the communication queue in the Task State by the SyncState component.

The system has a dedicated Message Dispatcher component that periodically checks each task's communication queue for undelivered messages.
It then delivers these messages to the recipient Agent specified in the message body.

The message dispatch by the Message Dispatcher triggers the receive message logic of the recipient Agent.
Based on whether the message attributes indicate a reply is needed, the recipient adds the corresponding processing step to its Agent Step:
\begin{itemize}
    \item If a reply is required, it adds a Send Message step.
    \item If no reply is needed, it adds a Process Message step.
\end{itemize}

Thus, a one-way, single-message sending process is shown in the lower half of Figure \ref{figure6}.
The sender generates a no-reply-needed message via Send Message.
After relay, the recipient passively adds a Process Message step to handle and understand the message content.

A two-way, long-term, multi-turn conversation process is shown in the upper half of Figure \ref{figure6}.
The sender generates a reply-requested message via Send Message.
Upon receiving it, the recipient adds a Send Message step to generate and send back a reply, while also deciding whether this new message requires further replies. 
\begin{itemize}
    \item If both Agents keep marking their latest messages as reply-required, they will continuously extend the dialogue via Send Message.
    \item If one Agent decides its latest message no longer needs a reply, it marks it as such in Send Message.
    The other Agent then ends the multi-turn dialogue with a Process Message step.
\end{itemize}

\section{Discussion}

\textbf{Agents Direct Communication}

It is particularly noteworthy that in our Allen architecture's MAS, both the initiation and termination of inter-agent communication are determined by Agents.
To be exact, it's by the Agent that created the most recent message.
An Agent can proactively add a 'Send Message' action to create message for other Agents.
Within the 'Send Message' capability, the Agent may also decide whether the message requires a reply.
The absence of a reply requirement typically signifies the end of that communication instance.

This closely mirrors human behavior: we may interrupt colleagues at will during work simply because we choose to, without even asking first.
Consequently, in Allen's architecture, prompt engineering must be carefully designed to prevent Agents from over-communicating about small things.
Each communication requires both participating Agents to temporarily allocate their threads to the communication steps (Send Message and Process Message), diverting resources from their primary tasks - particularly when step-locking mechanisms are triggered Appendix \ref{app:c}.

We also added another rule: Agents can only communicate if they're on the same Task.
Agents from completely different tasks can't message each other directly—all message must pass through the Task State.

\vspace{1em}
\textbf{Message Intervening Agent Actions}

Here we specify what forms of message-based intervention are allowed.
We require the Agent to autonomously generate Actions based on message content.
Specifically, we should not manually define how the Agent should respond to every non-command message, but instead allow it to make it's own decisions.

Thus, our message-handling logic leads to a decision branch that generates actions.
Specifically, it adds a skill step called 'Decision', where the Decision Step plans immediate short-term steps to interact with the MAS environment.

To achieve this, we introduce a new decison branch at both ends of the Agent's message-handling logic: the Send Message and Process Message skills.
These skills can now insert an additional Decision Step when needed to enable environment interaction, going beyond simply replying to or understanding messages.

\vspace{1em}
\textbf{Agent processes multi-stage tasks concurrently}

Our system does not restrict Agents from accepting tasks across multiple stages simultaneously.
However, since the Agent executes Steps in a single-threaded sequential manner, handling multi-stage task does not improve overall efficiency.
Instead, it may lead to interleaved execution of Steps from different stages, potentially prolonging each stages's completion time.

In terms of task focus, Agents closely mimic human work patterns. We should keep an Agent focused on a single task stage whenever possible.
To prevent Agents from handing too many stages concurrently, we can instantiate multiple functionally similar Agents to achieve true parallelism.

\vspace{1em}
\textbf{From Single-Agent System to Multi-Agent System}

From the perspective of policy autonomy, when our decision granularity shifts from workflow-level to agent-level, a single-Agent system naturally evolves into a multi-Agent system.
At this point, you gain the ability to select different Agents based on specific needs - where each Agent embodies a distinct workflow.

\vspace{1em}
\textbf{Allen enables simple iteration and optimization}

Our framework can accommodate any model-level optimizations, context-level improvements, or tool-level methods, as the July 2025 survey\cite{8} pointed out for iteration directions.
Our architectural innovation lies in redefining the operation mode of Multi-Agent Systems, while all model policies, model experiences, context prompts, context memories, and other module improvements can be freely replaced to adapt to the latest research approaches.
As our MAS system is driven by LLMs, its further refinement particularly benefits from agent system optimization methods based on LLM fine-tuning, such as \cite{9,10}

\section{Conclusion}
We introduce the Allen framework, a novel Multi-Agent System.
Within this framework, we propose using 'Step' as the minimal execution unit, enabling agents to iteratively execute their workflows step-by-step.
This approach can generalize to most scenarios with unprecedented policy autonomy, only requiring interaction with the system’s manager agent and no code modifications.

Furthermore, the Allen framework implements a step-wise execution paradigm.
Using four-level states to define and track complex tasks and multi-agent collaboration processes, it effectively balances collaboration efficiency, progress visibility, and human intervention capability.

\section*{Acknowledgments}
We acknowledge the support provided by Motern AI.

\section*{Author Contributions}
\textbf{Qiangong Zhou} : Core code contribution, architecture design.
\textbf{Zhiting Wang} : Architecture design, algorithm team leader.
\textbf{Mingyou Yao} : Frontend code contribution.
\textbf{Zongyang Liu} : Project supervision.

\bibliographystyle{unsrt}  
\bibliography{template}  

\appendix

\section*{Appendix}
For specific implementation details of the Allen architecture, please refer to the project's documentation (in Chinese) at: \url{https://github.com/motern88/Allen/blob/main/docs}

\section{Skills and Tools}
\label{app:a}
\textbf{Skill} refers to all LLM-driven derived capabilities. The key distinction lies in their prompt variations - each specific derived capability emerges based on different prompt configurations.

\textbf{Tool} represents capabilities that the LLM inherently lacks, but achieves through interfaces with external Agent modules. Compared to skills, tools are more aligned with real-world interactions, enabling the acquisition or modification of elements outside the Agent system.

\subsection{Skills}
All skills inherit common methods from the base executor class.

\vspace{1em}
\textbf{Planning}

The Agent uses the Planning skill to organize task execution steps, generating a multi-step execution plan.

Planning requires the ability to manipulate \texttt{AgentStep} within the Agent. \texttt{AgentStep} serves as the Agent's execution step manager, responsible for maintaining the Agent's step sequence. We employ prompt engineering to constrain the LLM to return planned step information in a specific format. Rule-based code then parses this information and adds the corresponding steps to \texttt{AgentStep}.

\vspace{1em}
\textbf{Reflection}

The Agent employs the Reflection skill to evaluate whether executed steps align with expectations. If adjustments are needed, it generates a new multi-step execution plan; otherwise, it appends a summarization step.

Reflection requires access to historical step execution data and the capability to append steps to \texttt{AgentStep}. We compile the results of past executions and phase objectives into a structured format for LLM processing. Through prompt constraints, the LLM returns reflection outcomes and planned step information in a predetermined format. Rule-based code subsequently parses this data and updates \texttt{AgentStep} accordingly.

Note: The Reflection skill cannot access stage information directly; phase objectives are obtained from Planning Step.

\vspace{1em}
\textbf{Summary}

The Agent uses the Summary skill to conclude and finalize a stage, marking its completion. It consolidates all step information within that stage and synchronizes it in the stage state.

The Summary skill cannot be planned during Planning—it is only planned when Reflection determines the task should end. Summary is solely responsible for aggregating the Agent's execution steps, not for delivering stage results. For example, if the stage goal is to output a text segment, the actual delivery process (e.g., via a tool like \texttt{send\_message}) should be handled by a dedicated delivery tool, not the Summary skill.

Summary requires access to historical step execution data. We compile the results of past executions and stage objectives into a structured format for LLM processing. Through prompt constraints, the LLM returns summarized outcomes in a predetermined format. Rule-based code then parses this data and synchronizes it to the Stage State.

Note: The Summary skill obtains stage information from the stage's initial step—the Planning step.

\vspace{1em}
\textbf{Instruction Generation}

This step generates the actual tool invocation instructions for the next tool step. Instruction Generation retrieves information about the next tool step and can update its content.

We input the next tool step's prompt information and instruction generation prompts into the LLM in a specific format, then capture the LLM's formatted instruction output. Rule-based code parses this information and updates the next tool step's instruction content accordingly.

\vspace{1em}
\textbf{Think}

The Agent utilizes the Think capability to handle text generation tasks requiring historical step information. This represents standard LLM-based text generation in the MAS that incorporates contextual data from previous steps.

\vspace{1em}
\textbf{Quick Think}

The Agent employs Quick Think for rapid response to text generation tasks that don't require historical step context. This constitutes a straightforward, single-instance LLM call/text generation within the MAS framework.

\vspace{1em}
\textbf{Send Message}

A unidirectional message sent by an Agent to another Agent instance within the MAS system.

Send Message first evaluates whether the current Agent's available information meets the message-sending criteria (i.e., whether the correct message content is known to the current Agent).
\begin{itemize}
    \item Information Retrieval Branch (activated when insufficient information exists):\\
    If the Agent lacks required information to send the message (determined by the LLM), it enters this branch to transform Send Message into a long-tail skill. A Decision Step is inserted to acquire additional information.
    \item Direct Message Branch (activated when sufficient information exists):\\
    Send Message aggregates all historical step execution data from the current stage, processes it via LLM according to the message intent, and delivers the message to the target Agent.
\end{itemize}

Implementation Details:
\begin{enumerate}
    \item Message Delivery Mechanism\\
    The message body is transmitted through the Executor's return result \texttt{execute\_output} (used to guide SyncState for state synchronization). \texttt{SyncState.sync\_state} places the message into the message processing queue of Task State. The MAS system's message processing module periodically scans this queue to execute message delivery tasks.
    
    \item Agent Communication Protocol/Flow\\
    Messages are processed by the receiver through appended steps (Process Message/Send Message):
    \begin{itemize}
        \item If the sender requires a reply: The receiver gets appended with a Send Message step directed to the sender.
        \item If no reply is needed: The receiver gets appended with a Process Message step (which doesn't involve message delivery or replies to other entities).
    \end{itemize}
    Thus:
    \begin{itemize}
        \item One-way messages are completed via Send Message → Process Message.
        \item Multi-turn dialogues are implemented through a series of Send Messages followed by one final Process Message.
    \end{itemize}

    \item Stage Affiliation Rules for Message Steps (Send Message/Process Message)\\
    \begin{itemize}
        \item If message delivery is task-phase-critical: Belongs to a stage. \\
        Ensures stage completion waits for message delivery.
        \item If message delivery is not task-phase-critical: Should not belong to any stage.\\
        \texttt{StepState.stage\_id} should be "no\_stage". Stage completion proceeds unaffected by message delivery status.
    \end{itemize}

    \item Message Waiting Mechanism \& Agent Step Locking
    \begin{itemize}
        \item When awaiting replies: Sender assigns unique wait IDs to all recipients (None if not waiting).
        \item When waiting: The initiating Agent won't execute any steps until all wait IDs are collected.
    \end{itemize}
\end{enumerate}

Agent-initiated Send Messages are typically task-phase-critical, thus require explicit \texttt{stage\_id} specification during sending.

\vspace{1em}
\textbf{Process Message}

The Agent processes a unidirectional message from another Agent instance within the MAS system, where the message explicitly requires no reply.

Upon receiving the message, the Agent uses the Process Message Step to invoke the LLM for handling the non-instructional portion of the message (the instructional part is processed by \texttt{agent\_base.process\_message}). Typically, this means the message needs to be digested and organized by the LLM, though it may also simply serve as the conclusion of a multi-turn dialogue.

Process Message understands and processes message content that requires no reply, while retaining the capability to react to message content through environmental interactions when necessary:
\begin{itemize}
    \item No Behavioral Response Required\\
    Digests and comprehends the message content, recording important parts into persistent memory.
    
    \item Behavioral Response Branch (Activated When Needed)\\
    The primary purpose of this branch is to enable Process Message to initiate environmental interaction behaviors. This is achieved by inserting and appending a Decision Step to plan short-term reactive behaviors.\\
    Workflow:
    \begin{enumerate}
        \item The LLM autonomously determines whether interaction behaviors are needed and returns corresponding instructions.
        \item Based on the LLM's returned instructions, we append and insert a Decision Step with matching attributes into the current Agent's step list
    \end{enumerate}
\end{itemize}

\vspace{1em}
\textbf{Task Message}

The Task Manager is a special skill (typically only available to manager-level Agents) that handles task management and scheduling.

The Task Manager uses its own historical step information (such as previously obtained task and stage details) to generate commands for managing task progress. The managing Agent then uses these commands to operate corresponding components in the MAS system, executing real actions. For example:
\begin{itemize}
    \item Using SyncState to update task and stage states.
    \item Sending messages to notify relevant Agents.
\end{itemize}

\vspace{1em}
\textbf{Agent Manager}

The Agent Manager is a special skill (usually restricted to manager-level Agents) that controls and coordinates other Agents.

The Agent Manager refers to its historical step data (such as previously gathered Agent information) to generate commands for directing other Agents. The managing Agent then uses these commands to operate the MAS system’s components, carrying out the required actions.

\vspace{1em}
\textbf{Ask Info}

The Ask Info skill allows an Agent to retrieve system/task details or information about other Agents.

Ask Info gives an Agent the ability to access external information, including other Agents' profiles and statuses.
\begin{itemize}
    \item SyncState helps collect higher-level data (e.g., stage state, task state).
    \item The retrieved information is sent back to the Agent via Message.
\end{itemize}

We use prompt engineering to guide the LLM to return specific commands in a structured format. These commands instruct SyncState to perform targeted queries, and the results are relayed back to the Agent through messages.

\vspace{1em}
\textbf{Tool Decision}

The Tool Decision skill allows an Agent to process the results of a long-tail tool and decide whether to continue or terminate its execution.

This skill invokes an LLM to handle the tool’s returned results and determines the next step—either guiding further tool usage (via instruction generation) or ending the call. If the tool’s execution continues, this skill appends an Instruction Generation step and another tool step for the Agent.

A long-tail tool call occurs when the tool’s results require repeated LLM verification and multiple executions. In such cases, the LLM iteratively decides the direction of each tool usage step.

After execution, the long-tail tool sends its results via SyncState (as a message), prompting the Agent to append a Tool Decision step to determine whether to continue or stop. Thus, the Tool Decision skill cannot be actively triggered by the Agent during Planning/Reflection—it is only activated by the long-tail tool.

Structure of a Long-Tail Tool Call:

Starts with Instruction Generation, ends with Too lDecision, and may include multiple (Instruction Generation → Tool Execution) cycles:
\begin{verbatim}
[I.G.] -> [Tool] -> [ToolDecision] -> [I.G.] -> [Tool] -> [ToolDecision] -> ...
\end{verbatim}

Notes:

\begin{itemize}
    \item The triggering of Tool Decision follows a standard cycle in MAS. Before executing this skill:\\
    Step (specific Tool execution) -> SyncState (generates command message) -> MessageDispatcher (distributes message to target Agent) -> Agent (processes message via \texttt{receive\_message}) -> Step (inserts a Tool Decision Step)
    
    \item After executing this skill, if Tool Decision continues the tool call:\\
    Step (Tool Decision skill confirms continuing tool call and appends subsequent steps) -> Step (Instruction Generation) -> Step (corresponding Tool execution)

    \item After executing this skill, if Tool Decision terminates the tool call:\\
    Step (Tool Decision skill terminates further tool calls)

    \item For MCP tool's long-tail calls, one key issue is how to pass the initially returned \texttt{capabilities\_list\_description} through Tool Decision to the next Instruction Generation.\\
    In \texttt{tool\_decision\_config.yaml} prompts, we specify in the tool step's \texttt{text\_content}:
    \begin{itemize}
        \item Detailed prompt text specifying the concrete objectives for the tool's next invocation.\\
        If you have obtained the capabilities\_list\_description returned by MCP Server, you should specify the exact invocation format of the MCP Server capability here (please write the complete dictionary of the returned description for the selected capability).
    \end{itemize}
\end{itemize}

\vspace{1em}
\textbf{Decision}

A more flexible, real-time decision-making skill. This skill is decoupled from Stage and does not rely on Stage state for decisions. Additionally, all steps planned by Decision are inserted (rather than appended at the end).

Key differences from other decision-making skills:
\begin{itemize}
    \item Stage-independent: Decision does not rely on Stage state, enabling more flexible responses to non-task-related decisions (e.g., replying to unexpected messages).
    \item Insert-based planning: Steps are inserted at priority positions rather than appended, giving Decision higher execution precedence.
\end{itemize}

\subsection{Tools}
All our tools strictly adhere to the Model Context Protocol (MCP) standard. This unified approach enables:

All our tools are implemented based on the Model Context Protocol (MCP) standard. Therefore, we only need to implement one tool executor—\texttt{MCPToolExecutor}—which inherits from the base Executor class. This eliminates the need to create a separate executor class for each MCP Server. However, for skills, we implement a dedicated \texttt{SkillExecutor} class for each type of skill.

Our tool implementation includes:
\begin{itemize}
    \item MCP Client: Provides basic MCP client functionality.
    \item MCP Tool Executor: The executor responsible for invoking specific MCP Server capabilities in the MAS system.
\end{itemize}

With this structure, we enable efficient scaling of any tool by loading various third-party MCP Servers into the MCP Client.

\vspace{1em}
\textbf{MCP Client Implementation}

We implement the MCP Client class to provide MCP-related functionalities to the Executor. In practice, we add additional caching and divide MCP connection management into four layers (where Layers 1, 3, and 4 are maintained by the \texttt{MCPClient} class, while Layer 2 corresponds to the actual tool permissions stored in the MAS \texttt{AgentState}).

\begin{itemize}
    \item Layer 1: \texttt{MCPClient.server\_config}\\
    Stores the startup configurations of all supported MCP Servers in the MAS.
    
    \item Layer 2: \texttt{AgentState.tools}\\
    Contains the permissions for external tools (MCP services) that the agent is allowed to call. The available MCP services in this layer are a subset of those in Layer 1.
    
    \item Layer 3: \texttt{MCPClient.server\_sessions}\\
    Maintains active MCP Server connection instances, where the key is the MCP Server name and the value is a `requests.Session` instance. The \texttt{server\_sessions} dynamically connect to the MCP Servers permitted in Layer 2, ensuring that all required MCP Servers (based on agent permissions) remain actively connected.
    
    \item Layer 4: \texttt{MCPClient.server\_descriptions}\\
    To avoid redundant overhead when fetching tool descriptions repeatedly, we also cache tool metadata. This layer stores detailed descriptions of available tools from the MCP Servers, where the key is the tool name and the value is its description. The \texttt{server\_descriptions} retrieves tool names, descriptions, and usage formats from the active sessions in Layer 3 and stores them.\\
    When an agent retrieves full tool and skill prompts, the \texttt{server\_descriptions} provides the necessary metadata. Similarly, when an agent executes a specific Tool Step or assembles a Tool Step prompt, the \texttt{server\_descriptions} supplies the relevant tool description and invocation format.
\end{itemize}

\vspace{1em}
\textbf{MCP Tool Executor Implementation}

The MCP Tool Executor enables agents to invoke any MCP (Model Context Protocol) server endpoint. This tool ensures compatibility with MCP-based server implementations in the MAS (Multi-Agent System).

From the agent's perspective, the MCP Tool connects to multiple MCP servers. However, from the MCP Server's viewpoint, this tool acts as its client.

Tool Execution Workflow in MAS: 
\begin{enumerate}
    \item Fetch MCP Server-Level Descriptions\\
    \begin{itemize}
        \item The agent retrieves predefined prompt descriptions for each tool server through its decision-making Skill Executor.
        \item Based on these descriptions, the agent decides whether to invoke the tool.
    \end{itemize}
    
    \item Fetch MCP Server Capability-Level Descriptions
    \begin{itemize}
        \item The MCP Tool Executor queries the MCP Client to obtain a list of all available capabilities supported by the target MCP Server.
        \item The agent then selects which specific capability to invoke.
    \end{itemize}

    \item Execute the Selected MCP Server Capability
    \begin{itemize}
        \item Based on the capability list from Step 2, the agent chooses a specific function to call.
        \item The MCP Tool Executor triggers execution via the \texttt{MCPClient.execute()} method, passing the capability name and input parameters, then returns the result
    \end{itemize} 
\end{enumerate}

During actual execution of the MCP Tool, both instruction generation and tool decision-making skills can access the basic MCP invocation prompts. It's important to note that while Agents already know how to invoke tools in MAS, they are not familiar with the MCP protocol.

The key purpose of these basic MCP invocation prompts is to provide Agents with visible interaction prompts related to MCP protocol execution, specifically:
\begin{enumerate}
    \item How to generate specific instructions for retrieving MCP Server capability lists, and how to interpret the returned capability lists;
    \item How to generate parameters for specific MCP Server capability calls, and how to understand the returned results from capability execution.
\end{enumerate}

\vspace{1em}
\textbf{Actual Invocation of MCP Client}

The MCP Client itself serves to manage connection sessions, and we intend to maintain only one globally unique MCP Client instance throughout the entire MAS. Given that our MAS architecture follows a multi-threaded parallel model with synchronous logic within each thread, we aim to enable concurrent execution of multiple MCP Client method calls within Agents (rather than sequential blocking). Our objectives are:
\begin{itemize}
    \item Enable parallel MCP operations within the same Agent
    \item Allow multiple Agents to share the same MCP Client event loop
\end{itemize} 

To achieve this, we need to implement a synchronous wrapper that internally uses \texttt{asyncio.run\_coroutine\_threadsafe} to submit tasks to the global event loop thread. This approach ensures:
\begin{itemize}
    \item When an Agent calls \texttt{MCPClient} → The entire system won't hang (only blocks the calling Agent thread)
    \item When multiple Agents call \texttt{MCPClient} → Concurrent execution occurs (since \texttt{MCPClient} runs in the event loop thread with asynchronous scheduling)
    \item Even when an Agent wants to initiate multiple concurrent MCP calls within a single Step, this can be achieved via \texttt{asyncio.gather} in the \texttt{MCPClient} event loop
\end{itemize} 

Consequently, we've separately implemented two key classes:
\begin{enumerate}
    \item \texttt{AsyncLoopThread} Class:\\
    \begin{itemize}
        \item Provides an asynchronous environment for the \texttt{MultiAgentSystem}
        \item Implements an asynchronous event loop thread for running async tasks in multi-threaded environments
        \item Enables Agents and Executors in MAS to submit async tasks to \texttt{AsyncLoopThread} without causing additional blocking
    \end{itemize} 
    
    \item \texttt{MCPClientWrapper} Class:\\
    \begin{itemize}
        \item Primarily used for invoking \texttt{MCPClient} methods within MAS
        \item Responsible for submitting \texttt{MCPClient} calls to the asynchronous event loop thread (\texttt{AsyncLoopThread})
        \item As a result, instead of passing \texttt{MCPClient} instances directly, MAS provides each Agent and tool Executor with an \texttt{MCPClientWrapper}
        \item Invoking \texttt{MCPClientWrapper} enables asynchronous calling of \texttt{MCPClient} methods within MAS
    \end{itemize} 
\end{enumerate}

\section{Persistent Memory}
\label{app:b}
In the Allen architecture's MAS, agent execution is divided into individual Steps. Each Step recomposes its prompts and does not share context with others. To address agents' long-term memory needs across Steps, Stages, and Tasks, we implemented the Persistent Memory mechanism.

\vspace{1em}
\textbf{Format of Persistent Memory}

We initialize a dictionary in Agent State (\texttt{agent\_state["persistent\_memory"] = {}}). During each Step execution, new memory entries are added with: 
\begin{itemize}
    \item Key: Timestamp (ISO format)
    \item Value: Memory content (natural language or structured data)\\
    \begin{verbatim}
    {
    "20250613T103022":"I've done...",
    "20250613T103523":"I'm doing...",
    "20250613T104023":"Recording task information...",
    }
    \end{verbatim}
\end{itemize}

\vspace{1em}
\textbf{Persistent Memory Management}

We have independently implemented a prompt that teaches the Agent to understand and manage persistent memory. This is integrated into the base Executor class used in every step, which includes methods to parse persistent memory instructions from LLM outputs and apply corresponding operations.

The Agent is allowed to manage its persistent memory using the following commands:
\begin{itemize}
    \item \textbf{Append new persistent memory}\\
    Use the 'add' command with new memory content. Append new persistent memory by adding the following formatted text to the output results:\\
    \begin{verbatim}
    <persistent_memory>
    [{"add":"Persistent memory content to append"}]
    </persistent_memory>
    \end{verbatim}
    
    \item \textbf{Delete existing memory entries}\\
    Use the 'delete' command with corresponding timestamps to remove specific memory entries. Delete/modify existing persistent memory by adding the following formatted text to the output results:\\
    \begin{verbatim}
    <persistent_memory>
    [{"delete":"Timestamps for permanent memory deletion"}]
    </persistent_memory>
    \end{verbatim}
\end{itemize}

\section{Communication Step Lock}
\label{app:c}

In our MAS, agents communicate with other components or agents through Send Message and Process Message skills executed at the Step level. To prevent dependent Steps (those requiring responses from pending Send Message operations) from executing prematurely and causing failures, we introduce the StepLock mechanism. 

The StepLock mechanism can suspend an agent's Step execution while awaiting critical message responses. Execution only resumes after all StepLocks are released (i.e., when all required message responses are successfully received). This ensures the logical dependencies between communication-dependent Steps and other Steps remain intact.

The StepLock mechanism in communication scenarios involves three core components: message reception logic (\texttt{AgentBase.received}) and Agent State, Send Message, and system message protocol (\texttt{Message}). We will systematically examine our key design consideration and the StepLock operation within those components.

\vspace{1em}
\textbf{Message}

As the standard message format for inter-agent communication in our MAS, the Message dictionary includes two dedicated fields for StepLock coordination:
\begin{enumerate}
    \item \texttt{waiting (Optional[List[str]])}\\
    (Sender-specified) Contains unique wait IDs corresponding to each receiver in \texttt{List[str]}\\
    \begin{itemize}
        \item When the sender requires response(s): Generates a unique wait ID per receiver
        \item Blocks all subsequent Steps until all wait IDs are released
        \item Default: None (no waiting required)
    \end{itemize} 
    
    \item \texttt{return\_waiting\_id (Optional[str])}\\
    (Receiver-specified in response) Echoes the sender's original wait ID\\
    \begin{itemize}
        \item Mandatory inclusion when responding to messages with waiting IDs
        \item Enables sender to release the corresponding wait lock
        \item Default: None (no ID to return)
    \end{itemize} 
\end{enumerate}

\vspace{1em}
\textbf{Send Message}

\begin{enumerate}
    \item When the Send Message skill calls the LLM to generate a preliminary message body, the LLM only needs to determine whether to reply and whether to wait.
    
    \item After the Send Message skill parses the LLM output, it will determine if the LLM considers the current message needs waiting.\\
    If waiting is required, it will automatically generate a unique waiting ID for each receiver in the preliminary message body, replacing the \texttt{["waiting"]} in the LLM-generated preliminary message body, and add the unique waiting ID to \texttt{agent\_state["step\_lock"]}.\\
    At this point, the \texttt{message["waiting"]} field value has changed from a boolean bool to an optional list containing unique waiting IDs for each receiver \texttt{optional[list[str]]}.
    
    \item When the Send Message skill constructs the \texttt{execute\_output}, it converts the LLM-generated preliminary message body into the MAS general message format Message:\\
    At this time, it constructs the \texttt{return\_waiting\_id} field value by attempting to extract it from \texttt{step\_state.text\_content}.\\
    (Generally, if an Agent receives a message with a \texttt{message["waiting"]} field value, when adding a reply step Send Message Step, it will extract the corresponding unique waiting ID that needs to be returned and place it at the end of the step's \texttt{text\_content}.)
\end{enumerate}

\end{document}